\documentstyle[preprint,aps]{revtex}
\begin{document}
\preprint{}
\draft
\title{\bf COLLIDING PLANE WAVES IN EINSTEIN-MAXWELL-DILATON FIELDS}
\author{Nora Bret\'on, Tonatiuh Matos and Alberto Garc\'{\i}a}
\address{Department of Physics\\ CINVESTAV, A.P. 14-740, 07000,MEXICO,D.F.}
\maketitle

\begin{abstract}

Within the metric structure endowed with two orthogonal space-like Killing
vectors a class of solutions of the Einstein-Maxwell-Dilaton field equations
is presented. Two explicitly given sub-classes of solutions bear an
interpretation as colliding plane waves in the low-energy limit of the
heterotic string theory.

\end{abstract}
\vspace{.3in}

\hspace{.35in} PACS number(s) : 04.50.th, 04.40.Nr, 11.25.Mj
\newpage

{\bf 1. Introduction}

The study of the gravitational interaction coupled to the Maxwell and dilaton
fields has been the subject of recent investigations related to the heterotic
string theory.  Dilaton fields coupled to Einstein-Maxwell fields appear in a
natural manner in the low-energy effective action in string theory and as a
result of a dimensional reduction of the Kaluza-Klein Lagrangian. It has been
realized that the low-energy effective field, which describes string theory,
contains solutions endowed with qualitatively different features from those
ones that appear in ordinary Einstein gravity [1].

Lately it has been found that plane wave geometries are exact solutions for
the string theory to all orders of string tension parameter [2]. It is
therefore of interest to consider the collision of plane gravitational waves
with electromagnetic and dilaton fields. In fact, some solutions of this kind
have been already presented by G\"urses [3].

In the context of General Relativity the topic of colliding plane
gravitational waves has been widely explored and colliding wave solutions with
scalar fields have been found too. However, those scalar fields were weakly
coupled to the electromagnetic field [4], while the most intriguing features
of string gravity are due to the peculiar nature of the dilaton heterotic
coupling to vector fields. Here we consider the stringy gravity model
including vector fields for colliding plane gravitational waves, i. e., the
Einstein-Maxwell-Dilaton (EMD) system with an arbitrary dilaton coupling
constant in the framework of interacting plane waves.

We consider the action [5]

\begin{equation}
S= \int{d^4x \sqrt{-g} \{-R+2(\nabla \Phi)^2 +e^{-2 \alpha \Phi} F^2 \}},
\end{equation}
where $g={\rm det}(g_{\mu \nu}), \mu, \nu = 0, 1, 2,3$. $R$ is the scalar
curvature, $F_{\mu \nu}$ is the Maxwell field, and $\Phi$ is the dilaton
field. The constant $\alpha$ is a free parameter which governs the strength of
the coupling of the dilaton to the Maxwell field. Special theories are
contained in (1): For $\alpha =\sqrt{3}$, the action (1) leads to the Kaluza-
Klein field equations obtained from the dimensional reduction of the five-
dimensional Einstein vacuum equations. For $\alpha=1$, the action  (1)
coincides with the low energy limit of string theory with vanishing dilaton
potential [6]. Finally, in the extreme limit $\alpha=0$, (1) yields the
Einstein-Maxwell theory minimally coupled to the scalar field.

The field equations obtained from (1) are

\begin{equation}
(e^{-2 \alpha \Phi} F^{\mu \nu})_{; \mu}=0,
\end{equation}
\begin{equation}
\Phi^{; \mu}_{; \mu} + {\alpha \over 2}e^{ -2 \alpha \Phi}F_{\mu \nu}F^{\mu
\nu} =0, \end{equation}

\begin{equation}
R_{\mu \nu} = 2 \Phi_{, \mu} \Phi_{, \nu} +2 e^{-2 \alpha \Phi} (F_{\mu
\lambda} F_{\nu}^{\lambda} -{1 \over 4}g_{\mu \nu} F_{\alpha \beta}F^{\alpha
\beta}), \end{equation}
where a semicolon denotes the covariant derivative with respect to $g_{\mu
\nu}$. A few exact solutions of Eqs. (2)-(4) are known; they reveal many
interesting features of the dilaton field (see [1] and references therein). In
this paper we present solutions to Eqs. (2)-(4) with a colliding plane wave
interpretation. We first present the solutions in the interaction region and
then extend them beyond the null boundaries. In the next section we outline
the usual representation of the colliding plane wave spacetime in General
Relativity and the corresponding field equations. In section 3 we present
explicitly the solutions and check that the appropriate boundary conditions
for colliding waves are satisfied. In section 4 we comment about the nature of
the singularity and finaly we draw some conclusions in section 5.

\bigskip

{\bf 2.  The Colliding Waves Spacetime and the Field Equations }

A spacetime describing the collision of plane waves admits two spacelike
Killing vector fields. In this work we take them to be orthogonal. For such a
case, we consider the metric $g_{\mu \nu}$ and the U(1) gauge potential
$A_{\mu}$ as given by

\begin{equation}
ds^2= 2 e^{-M} du dv +e^{-U} (e^{-V} dy^2+e^V dx^2),
\end{equation}
\begin{equation}
A_{\mu}=(0,0,A,0),
\end{equation}
where $M=M(u,v), U=U(u,v), V=V(u,v), A=A(u,v)$ and the electromagnetic field
is $F_{\mu \nu} =  A_{\nu, \mu} - A_{\mu , \nu}$.

The spacetime for the collision of plane waves is divided into four disjoint
regions: Region I (of interaction): $0 \le u \le 1, \quad 0 \le v \le 1$.
Region II : $u<0, \quad 0<v<1;$ and Region III : $0<u<1, \quad v<0,$ where
``live" the incoming waves. The boundaries between the region I and regions II
and III are $u=0$ and  $v=0$. Finally, it is considered the region IV : $u<0,
\quad v<0$, which corresponds to the spacetime before the pass of any wave.
The line element (5) applies to the entire spacetime, however the metric
functions $U, V$ and $M$ must take different forms in the four regions.

 The field equations (2)-(4) turn out to be

\begin{equation}
-2 A_{,uv}= (V_{,u}- \alpha \Phi_{,u})A_{,v}+
(V_{,v}- \alpha \Phi_{,v})A_{,u} \quad ,
\end{equation}

\begin{equation}
U_{,uv}=U_{,u} U_{,v} \quad ,
\end{equation}

\begin{equation}
2M_{,uv}=-2U_{,uv}+U_{,u}U_{,v}+V_{,u}V_{,v}+4 \Phi_{,u} \Phi_{,v}  \quad ,
\end{equation}

\begin{equation}
2V_{,uv}-U_{,u}V_{,v}-U_{,v}V_{,u}-4e^{U+V- \alpha \Phi} A_{,u}A_{,v}=0,
\end{equation}

\begin{equation}
2\Phi_{,uv}-U_{,u} \Phi_{,v} -U_{,v} \Phi_{,u} +{\alpha \over 2} e^{U+V-
\alpha \Phi} A_{,u}A_{,v}=0, \end{equation}

\begin{equation}
-2 M_{,u}U_{,u}-2U_{,uu}+U_{,u}^2+V_{,u}^2+4 \Phi_{,u}^2+4e^{U+V- \alpha \Phi}
A_{,u}^2=0, \end{equation}

\begin{equation}
-2 M_{,v}U_{,v}-2U_{,vv}+U_{,v}^2+V_{,v}^2+4 \Phi_{,v}^2+4e^{U+V- \alpha \Phi}
A_{,v}^2=0, \end{equation}

The dilaton field $\Phi= \Phi(u,v)$. Note that Eq. (9) can be derived from the
other equations. Eq. (8) can be immediately integrated
\begin{equation} e^{-U}= a(u) + b(v), \end{equation}
with $a$ and $b$ being  arbitrary functions of $u$ and $v$ respectively.

The corresponding components of the Weyl tensor are computed to be

\begin{equation}
\Psi_o^o=-{1 \over 2} [ V_{,vv}-V_{,v}(U_{,v}-M_{,v})],
\end{equation}

\begin{equation}
\Psi_4^o=-{1 \over 2} [ V_{,uu}-V_{,u}(U_{,u} -M_{,u})],
\end{equation}
\begin{equation}
\Psi_2^o={1 \over 2} M_{,uv}, \quad \Psi_1^o=\Psi_3^o=0,
\end{equation}
We shall give in the next section the solution for region I and then we
discuss the  matching to the precolliding regions.

\bigskip

{\bf 3. The EMD solutions}

Although one can proceed with the above $(u,v)$-dependence formulation, it
occurs more effective -from the integration point of view- to use a
$(\rho,z)$-dependence, i. e., to look for solutions for the EMD Eqs. (2)-(4)
for a diagonal line element of the form

\begin{equation}
ds^2= {e^{2k} \over f}(d \rho ^2 - dz^2) + \rho [ \rho f^{-1} dx^2+ \rho^{-1}
f dy^2], \end{equation}
with $\partial_x$ and $\partial_y$ being the two commuting spacelike Killing
vectors, and $f, k$ being functions of $\rho$ and $z$ only. We can arrive to
(18) from the metric (5) by defining
\begin{eqnarray} \rho&=& e^{-U}= a(u)+ b(v), \nonumber\\ z&=&a(u)-b(v),
\end{eqnarray}
and identifying $2k \to -(M+V+U)- \ln{[2 a'(u)b'(v)]}$ and $f \to \exp{[-
(V+U)]},$ where $a'(u)$ and $b'(v)$ denote the derivatives in $u$ and $v$
respectively.

The method used to determine  the sought solutions is the harmonic mapping
combined with the algebra associated to the group $SL(2, {\bf R})$, which
reduce the integration of the Einstein's equations to an algebraic problem
(see [7] and references therein). Other methods to obtain solutions have been
addressed like the inverse scattering method [8], however, we encounter that
by means of the harmonic map one gets wider class of solutions in a more
straightforward manner.

A class of solutions for the EMD Eqs. (2)-(4) is given by
\begin{equation}
f={f_o e^{\lambda} \over (a_1 \Sigma_1 +a_2 \Sigma_2)^{\gamma}},
\end{equation}
\begin{equation}
\kappa^2 = e^{-2 \alpha \Phi} = \kappa_o^2 (a_1 \Sigma_1 +a_2
\Sigma_2)^{\beta}e^{ \lambda - \tau_o \tau},
\end{equation}
\begin{equation}
A = A_{y} ={ {(a_3 \Sigma_1 +a_4 \Sigma_2)} \over {(a_1 \Sigma_1 +a_2
\Sigma_2)}},
\end{equation}
where $\Sigma_1$ and $\Sigma_2$ denote functions on the variable
$\tau(\rho,z)$ which is determined by the harmonic map (Eq.(14) in Ref. [7]);
for each pair $(\Sigma_1, \Sigma_2)$ we have a different solution for the Eqs.
(20)-(22) (See Eqs.(25) and (27) below); $ \tau_o, \kappa_o, f_o, a_1, a_2,
a_3,$ and  $a_4$ are constants, and $\gamma$ and $\beta$ are $\alpha$-
dependent parameters

\begin{equation}
\gamma= {2 \over{1+ \alpha^2}}  , \quad \beta= {2 \alpha^2 \over {1+
\alpha^2}}.
\end{equation}
The functions $ \lambda( \rho,z)$ and $\tau (\rho, z)$ are each one a solution
of the equation
\begin{equation}
\phi_{, \rho \rho} +{1 \over \rho} \phi_{, \rho} - \phi_{,zz}=0. \end{equation}
Among the above solutions (20)-(22), we distinguish two cases

case $(i)$

\begin{equation}
\Sigma_1=e^{q_1 \tau}, \quad \Sigma_2=e^{q_2 \tau}, \quad 4 a_1a_2f_o +
\kappa_o^2(1+ \alpha^2)(a_1a_4-a_3a_2)^2=0,
\end{equation}
where $q_1$ and  $q_2$ are constants. The corresponding equations for $k,$ the
transversal gravitational degree of freedom, are

\begin{eqnarray}
k_{,z}&=& {\rho \over 2} \{ ({{\alpha^2+1} \over \alpha^2})\lambda_{,\rho}
\lambda_{,z} - (2 \gamma q_1q_2 - { \tau_o^2 \over \alpha^2}) \tau_{,\rho}
\tau_{,z} -{\tau_o \over \alpha^2}( \tau_{,z} \lambda_{,\rho} + \tau_{,\rho}
\lambda_{,z}) \}, \nonumber\\ k_{,\rho}&=& {\rho \over 4} \{ ({{\alpha^2+1}
\over \alpha^2})(\lambda_{,\rho}^2+ \lambda_{,z}^2) - (2 \gamma q_1q_2 - {
\tau_o^2 \over \alpha^2}) (\tau_{,\rho}^2+ \tau_{,z}^2) -{2 \tau_o \over
\alpha^2}( \tau_{,z} \lambda_{,z} - \tau_{,\rho} \lambda_{,\rho}) \},
\end{eqnarray}
which are integrable once one specifies $\lambda (\rho,z)$ and
$\tau(\rho,z)$, solutions of Eq. (24).

case $(ii)$

\begin{equation}
\Sigma_1= \tau, \quad \Sigma_2= 1, \quad q_1=- q_2, \quad 4 a_1^2f_o -
\kappa_o^2(1+ \alpha^2)(a_1a_4-a_3a_2)^2=0,
\end{equation}
The corresponding equations for $k$ are
\begin{eqnarray}
k_{,z}&=& \rho \lambda_{,\rho} \lambda_{,z} , \nonumber\\
k_{,\rho}&=& {\rho \over 4} (\lambda_{,\rho}^2+ \lambda_{,z}^2) .
\end{eqnarray}
which, again,  are integrable as soon as one specifies $\lambda (\rho, z)$,
solution of Eq. (24).

The solutions of Eq. (24) are of the form
\begin{eqnarray}
\phi &= & K {\rm ln} \rho +L \{ A_{\omega} \cos [\omega (z+z_o)]J_o(\omega
\rho) \}+ \nonumber\\ & &L \{ B_{\omega} \cos [\omega (z+z_o)]N_o(\omega \rho)
\} - \sum_{i}d_i {\rm arccosh} ({{z+z_i} \over \rho}),
\end{eqnarray}
where $K$ is a constant, $L \{ \}$ stands for arbitrary linear combinations of
the terms in curly brackets and $J_o(\omega \rho)$ and $N_o(\omega \rho)$  are
the Bessel and Neumann functions of zero order respectively.

An explicit relationship between the coordinates $(\rho,z)$ of the metric (18)
and the null coordinates $(u,v)$ of the metric  (5) is given when we select
$a(u)= {1 \over 2} - u^n, \quad b(v)= {1 \over 2} - v^m,$, then we have

\begin{equation}
\rho= 1-u^n-v^m, \quad z=v^m-u^n,
\end{equation}
with $m$ and $n$ being constants determined by boundary conditions. The null
coordinates $(u,v)$ are more suitable for the analysis of the matching
conditions, which we address in the next subsection.

{\bf Continuity of the Metric on the Null Boundary}

The solutions for cases $(i)$  and $(ii)$ can be interpreted as the
gravitational field in the interaction region arising after the collision of
two gravitational plane waves only if certain boundary conditions on the null
hypersurfaces $u=0$ and $v=0$ are satisfied [9]. With the chosen coordinate
relation, Eq.(30), one has to verify only the continuity on $u=0$ and $v=0$ of
the metric coefficient $g_{uv}=4mn u^{n-1}v^{m-1} {e^{2k} f^{-1}}$, which
arises when we substitute Eq. (30) in (18), and taking the expression for $f$,
Eq. (20), one arrives at

\begin{equation}
g_{uv}= 4 m n u^{n-1}v^{m-1}{e^{2k}f_o^{-1}} \{a_1 \Sigma_1 +a_2 \Sigma_2
\}^{\gamma} e^{- \lambda}. \end{equation}

We shall prove separately the continuity on $u=0$ and $v=0$ of the appearing
above factors

\begin{equation}
(a_1 \Sigma_1+a_2 \Sigma_2)^{\gamma}e^{ - \lambda},
\end{equation}
\begin{equation}
{\rm and} \quad u^{n-1}v^{m-1}e^{2k}.
\end{equation}

For the case $(i)$, without loss of generality, we can take as solutions for
$\tau$ and $\lambda$ the following functions [10]

\begin{eqnarray}
\tau &= & d_1 {\rm arccosh}[{z+1 \over \rho} ] = d_1 \ln [{z+1 \pm
\sqrt{(z+1)^2- \rho^2} \over \rho}] , \nonumber \\ \lambda &= & d_2 {\rm
arccosh} [{1-z \over \rho}]= d_2 \ln [{1-z \pm \sqrt{(1-z)^2- \rho^2} \over
\rho}],
\end{eqnarray}
where $d_1$ and $d_2$ are constants. Substituting the expressions (34) into
(32) and taking separately the limits $u \to 0$ and $v \to 0$ (noting that
$u=0$ corresponds to $\rho=-z+1$ while $v=0$ corresponds to $\rho=z+1$), it is
easy to see that the factor (32) does not diverge on $u=0$ neither on $v=0$.
Thus we are led with the factor (33), i. e., $u^{n-1}v^{m-1}e^{2k}.$ To ensure
the smooth matching between the interaction and the precollision regions, the
function $e^{2k}$ must diverge as $u^{1-n}$ and $v^{1-m}$ on $u=0$ and $v=0$
respectively. This divergence in $e^{2k}$ comes from the terms of Eqs. (34);
to show that, we note from Eqs. (26) that one can split the function $k$ as

\begin{equation}
k={{\alpha^2+1} \over {2 \alpha^2}}k_g-(\gamma q_1 q_2 - {\tau_o^2 \over {2
\alpha^2}} )k_e - {\tau_o \over {2 \alpha^2}} k_s,
\end{equation}
consequently

\begin{equation}
e^{2k}= e^{({{\alpha^2+1} \over {2 \alpha^2}})2k_g} e^{-{(\gamma q_1 q_2 -
{\tau_o^2 \over {2 \alpha^2} })}2k_e} e^{ -( {\tau_o \over {2 \alpha^2}})2
k_s} \equiv  e^{2K_2 k_g} e^{2K_1 k_e} e^{2K_3 k_s},
\end{equation}
where $k_g, k_e$ and $k_s$ are solutions of the following set of equations
\begin{eqnarray}
k_{g,z}&= & \rho \lambda_{,\rho} \lambda_{,z}, \nonumber\\
k_{g,\rho} &= & {\rho \over 2}  (\lambda_{,\rho}^2 + \lambda_{,z}^2),
\end{eqnarray}

\begin{eqnarray}
k_{e,z}&= & \rho \tau_{,\rho} \tau_{,z}, \nonumber\\
k_{e,\rho} &= & {\rho \over 2}  (\tau_{,\rho}^2 + \tau_{,z}^2),
\end{eqnarray}

\begin{eqnarray}
k_{s,z}&= & \rho ( \tau_{,z}  \lambda_{,\rho} +\tau_{,\rho} \lambda_{,z}),
\nonumber\\ k_{s,\rho}&= & \rho ( \tau_{,z}  \lambda_{,z} +\tau_{,\rho}
\lambda_{,\rho}),
\end{eqnarray}

Integrating Eqs. (39) with $\lambda$ and $\tau$ given by Eqs. (34), it turns
out that the factor $ e^{2K_3 k_s} $  does not diverge neither on $u=0$ or
$v=0$. Furthermore, performing an analogous analysis as in [10], it can be
shown that $\tau$ contributes to the function $k_e$, via Eqs. (38), with the
following term on $v=0$,

$$-{1 \over 2} d_1^2 \ln{[(z+1)^2- \rho^2]}=-{1 \over 2} d_1^2 \ln{(v^m)}+
{\rm bounded} \quad {\rm terms}, $$ which gives the desired behaviour if $K_1
d_1^2 = 2 - {2 \over m}$. Analogously, $\lambda$ contributes to the function
$k_g$, via Eqs. (37), with the term on $u=0$ of the form

$$-{1 \over 2} d_2^2 \ln{[(1-z)^2- \rho^2]}=-{1 \over 2} d_2^2 \ln{(u^n)}+
{\rm bounded} \quad  {\rm terms}$$ which behaves properly if $K_2 d_2^2 = 2 -
{2 \over n}.$

We can use solutions for $\lambda$ and $\tau$ as those given by Eq. (29)
involving more terms; however,  all other   contributions of $\lambda$ and
$\tau$ to the function $e^{2k}$ are found to be bounded on $u=0, v=0$.
Therefore, provided there exist at least two terms  of the form given by  Eqs.
(34), in the case (i) the verification of the boundary conditions relevant to
the colliding wave problem is ensured if the constants fulfill the conditions

\begin{equation}
K_1 d_1^2 = 2 - {2 \over m}, \qquad K_2 d_2^2 = 2 - {2 \over n}.
\end{equation}
For the case $(ii)$ the previous analysis apply when one chooses

\begin{equation}
\lambda =  c_1 {\rm arccosh}[{z+1 \over \rho}] + c_2 {\rm arccosh}[{1-z \over
\rho}],
\end{equation}
and $\tau$, for instance, can be chosen as in (34). Again, it can be shown
that the term (32) does not diverge neither on $u=0$ or $v=0$. In relation to
the term (33), the constants $c_1$ and $c_2$ can be adjusted conveniently  in
order to achieve an smooth matching of the solution on $u=0$ and $v=0$.  The
previous analysis showed that the solutions given by Eqs. (25)-(26) subjected
to (40) and (27)-(28) can be interpreted as colliding wave fields.

{\bf Behaviour of the Fields on the Null Boundaries}

{}From Eq. (22) the nonvanishing components of the electromagnetic field turn
out to be

\begin{equation}
F_{y \rho}={a_1a_4-a_3a_2 \over{(a_1 \Sigma_1 +a_2 \Sigma_2)^2}} \{\Sigma_2
\Sigma_{1, \rho}- \Sigma_1 \Sigma_{2, \rho} \}, \end{equation}

\begin{equation}
F_{y z}={a_1a_4-a_3a_2 \over{(a_1 \Sigma_1 +a_2 \Sigma_2)^2}} \{\Sigma_2
\Sigma_{1, z}- \Sigma_1 \Sigma_{2, z} \},
\end{equation}
if we choose, for example, $\tau$ as in Eq.(34) it is straightforward to show,
from Eq. (25) for case (i) and from Eq.(27) for case (ii), that $F_{\mu \nu}$
does not diverge on $u=0$ neither on $v=0$.

For the dilaton field $\Phi$ ,in the case (i), substituting the Eqs.(25) and
(34) in Eq. (21) and taking separately the limits $u \to 0$ ($\rho \to 1-z$)
and $v \to 0$ ($\rho \to 1+z$)  it can be shown that $\kappa^2=e^{-2 \alpha
\Phi}$ does not diverge on $u=0$ neither on $v=0$ and this behaviour is
independent of the constants $\tau_o, d_1, d_2, q_1, q_2, \alpha $. The
analogous occurs for case (ii) subtituting in (21) the Eqs. (27) and (34). For
the precolliding region IV ($u \le 0, \quad v \le 0$), for the case (i), the
dilaton field becomes a constant, $\kappa^2 = \kappa_o^2$, while for the case
(ii) the value of $\kappa$ vanishes.

\bigskip

{\bf 4 Singularities and Discontinuities of the Curvature}

{\bf along the Null  Boundaries}

We now discuss briefly the behaviour of the fields on the null boundaries
$u=0$ and $v=0$ in the context of the field equations. In order to do this we
pass from region I to region II and III using the Penrose's procedure [11] :
the  continuations of the fields from region I to the remaining regions II and
III, and further to IV can be achieved by replacing the coordinates $u$ and
$v$ in accordance with

\begin{equation}
u \to uH(u), \qquad v \to vH(v),
\end{equation}

As a consequence of this procedure, singularities or discontinuities (or both)
of the Riemann tensor can arise on the null hypersurfaces. To determine their
behaviour we follow the analysis accomplished by Chandrasekhar and
Xanthopoulos [12]. In their paper they showed that the quantities involving
first derivatives of the metric functions can at most suffer an $\Theta$-
function discontinuity, while those quantities with second derivatives in the
coordinates $u$ or $v$, can involve $\delta$-function singularities. With this
criteria, we can characterize the behavior of the  fields, Einstein tensor
components and curvature on the null boundaries.

{}From the field Eqs. (7)-(13) we can see that  they involve first derivatives
and terms of the form of mixed derivatives $ \partial^2 / \partial_u
\partial_v $, but mixed derivatives do not lead to $\delta$-function
distributions, then the fields are consistent on the null boundaries, provided
we select $U_{,uu}$ and $U_{,vv}$ such that this second derivative do not lead
to a $\delta$-function behavior. The curvature components (15)-(17) behave as
it should be on the null boundaries; for a detailed general analysis in this
respect see [13]

\bigskip

{\bf Singularities on the Focussing Hypersurface}

Colliding plane wave solutions exhiibit singularities at the so called
foccusing surface. The origin of this singularity has been discussed in [14].
{}From  the metric (18), we realize that singularities can arise when $f=0$ or
if $e^{2k}$ diverges. Both behaviours can occur when $\rho= 1-v^m-u^n=0$. For
the case (i), we balance separately each term by arranging the constants
properly. For $f$ we have ( writting only the terms which depend on $\rho$):

\begin{eqnarray}
& &e^{-d_2 \ln \{ \}}(a_1e^{q_1d_1 \ln \{ \}}+a_2e^{q_2d_1 \ln \{
\}})^{\gamma} \nonumber\\ &  & \simeq  \big[ {{1-z \pm \sqrt{(1-z)^2 -
\rho^2}} \over \rho}\big]^{-d_2} \{ a_1 \big[ {{z+1 \pm \sqrt{(z+1)^2 -
\rho^2}} \over \rho} \big]^{q_1 d_1}+ \nonumber\\ && a_2 \big[ {{1+z \pm
\sqrt{(1+z)^2 - \rho^2}} \over \rho}\big]^{q_2d_1} \} \nonumber\\ & & \simeq
\rho^{d_2-q_1d_1 \gamma} [1+() \rho^{-q_2d_1+q_1d_1}]^{\gamma}
\end{eqnarray}
expanding the term in brackets, we take the highest power in $\rho$, $\rho^{(-
q_2d_1+q_1d_1)\gamma}$. This term must balance the term outside the bracket.
If the constants can be adjusted in such a manner that $d_2-q_1d_1 \gamma +
\gamma(-q_2d_1+q_1d_1) >0$, or $d_2 > q_2 d_1 \gamma$, then this term does not
diverge at $\rho=0$. Examining now the factor $e^{2k}$, from the integration
of Eqs. (37)-(39) the terms which diverge at $\rho=0$ are $$e^{2k} \simeq
\rho^{K_2d_2^2+K_1d_1^2+2K_3d_1d_2}({\rm bounded \quad terms})$$ Therefore,
the singularity will be avoided if we impose one more  condition :

\begin{equation}
K_2d_2^2+K_1d_1^2+2K_3d_1d_2 \ge 0.
\end{equation}
Consequently, imposing on the constants the conditions determined above, the
singularity can be avoided for the case (i).

For the case (ii) the term corresponding to $e^{2k}$ can always be arranged to
be not divergent, however, for $f$ we have $$e^{- \lambda}(a_1 \tau  +
a_2)^{\gamma} \simeq {\rm bounded} \quad {\rm terms} +() \rho^{c_1+c_2}
(\ln{\rho})^{a_1d_1\gamma},$$ the last term can not be balanced with the
another term, thus in this case the singularity can not be avoided.

{\bf 5 Final Remarks }

In this paper it is considered the problem of the field arising as a result of
collision of  plane gravitational waves in the Einstein-Maxwell-Dilaton
fields. Two solutions of the Einstein-Maxwell-Dilaton equations interpretable
as colliding gravitational plane waves are explicitly given.  The metric is
diagonal,  this means that the two commuting Killing vectors are orthogonal.
The verification of the boundary conditions relevant to the colliding wave
problem are determined essentially by the physical structure of the incoming
plane waves, whose  ``amplitude" must be adjusted (Eqs. (40)) depending on the
values of the coupling constant $\alpha$, the constant of the dilaton field
$\tau_o$, and the constants $q_1,\quad  q_2$ of the metric functions. For the
boundary conditions the waves act separately on each boundary $u=0$ and $v=0$.
It is discussed briefly the behavior of the fields on the null boundaries. In
relation to the singularity developed after the waves collide,  it occurs that
to avoid the singularity, for our first case, it must be imposed conditions
which involve both  amplitudes and also the coupling constant $\alpha$. In
contrast, for the case (ii), the singularity cannot be avoided by tuning
properly the free parameters. Remains as an open question if the solutions
presented here can be extended to all orders in the string tension parameter.

\bigskip

{\bf Acknowledgements}

This work was partially supported by CONACyT (M\'exico).

\vfill\eject

{\bf References}
\begin{enumerate}
\item D. Garfinkle, G. Horowitz and A. Strominger, Phys. Rev. {\bf D 43}, 3140
(1991); G. Gibbons and K. Maeda,  Nucl. Phys. {\bf B298}, 741 (1988).
\item R. G\"uven, Phys. Lett. {\bf B191}, 275 (1987);
G. Horowitz and A. Steif, Phys. Rev. Lett. {\bf 64}, 260 (1990).
\item M. G\"urses and E. Sermutlu, Phys. Rev. {\bf D 52}, 809 (1995)
\item J. B. Griffiths, {\it Colliding Plane Waves in General Relativity},
(Oxford: Oxford Univ. Press), 1991, Ch. 20.
\item J. H. Horne and G. T. Horowitz, Phys. Rev. {\bf D46}, 1340 (1992).
\item J. H. Horne and G. T. Horowitz, Phys. Rev. {\bf D48}, R5457, (1993).
\item N. Bret\'on and T. Matos ``Colliding waves as harmonic maps" in {\it
Gravitation: The Spacetime Structure}, Proceedings of the 8th Latin American
Symposium on Relativity and Gravitation, Aguas de Lindoia, Brazil, 1993. ed.
by P. Letelier and W. Rodriguez (World Scientificc, Singapore, 1994)
\item S. Mizoguchi. ``Colliding Wave Solutions , Duality and Diagonal
Embedding of General Relativity in two-dimensional Heterotic String Theory",
PREPRINT DESY95-126.
\item P. Szekeres, J. Math. Phys. {\bf 13}, 286, (1972);
S. O'Brien and J. L. Synge, Proc. Dublin Inst. Adv. Stud. {\bf
A9}, 1 (1952).
\item A. Feinstein and J. Ib\'a\~nez, Phys. Rev. {\bf D39}, 470,
(1989).
\item  K. Khan and R. Penrose, Nature (London), {\bf 229}, 185 (1971).
\item S. Chandrasekhar and B. C. Xanthopoulos, Proc. R. Soc. Lond. {\bf A398},
223 (1985).
\item A. Garc\'{\i}a, Theor. and Math. Phys. {\bf 83}, 434 (1990).
\item R. A. Matzner and F. J. Tipler, Phys. Rev. {\bf D29}, 1575,
(1984).

\end{enumerate}
\end{document}